\documentclass[]{article}

\usepackage{amsmath}
\usepackage{amsfonts}
\usepackage{amssymb}
\usepackage{amsthm}
\usepackage{a4}
\usepackage{amssymb,amsmath}
\usepackage{graphicx}
\usepackage{epsf}
\usepackage{bbm}
\usepackage{bm}

\begin{document}

\title{The 2S$_{1/2}$--2P$_{1/2}$ Lamb Shift in He$^+$} 

\author{U.~D.~Jentschura${}^1$ and G.~W.~F.~Drake${}^2$\\[1ex]
{\em ${}^1$Universit\"at Freiburg, Physikalisches Institut (Theoretische 
Quantendynamik),}\\
{\em Hermann--Herder--Stra\ss{}e 3, 79104 Freiburg im 
Breisgau, Germany}\\
{\em ${}^2$Department of Physics, University of Windsor, Windsor, Ontario 
N9B 3P4}}

\maketitle 

\begin{abstract} 
The current theoretical status of the Lamb shift in He$^+$ is 
discussed. Recent calculations of two-loop binding corrections to the 
Lamb shift significantly shift the theoretical value of the ``classic'' 
Lamb shift in He$^+$, i.e.~of the 2S$_{1/2}$--2P$_{1/2}$-interval. In 
this brief research note, we present a new (theoretical) value for this 
interval which reads $14041.474(42)\,{\rm Mhz}$. The theoretical 
uncertainty is reduced as well as the discrepancy between theory and 
experiment. Planned measurements should be of help in further 
elucidating the situation. 
\end{abstract} 

%
%
\section{Introduction} 

One of the intriguing questions which remain to be answered in 
bound-state quantum electrodynamics is related to the discrepancy 
between the theoretical and experimental values for the Lamb shift in 
ionized helium, or He$^+$, which is a hydrogenlike atomic system with a 
nuclear charge number $Z=2$. The latest measurement of the 
``classical'' Lamb shift in He$^+$, using the anisotropy method, has 
led to the result~\cite{vWHoDr2001} 
\begin{equation} 
\label{nuexp} 
\nu_{\rm exp}(2{\rm S}_{1/2} - 2{\rm P}_{1/2}) = 14041.13(17) \, {\rm 
MHz}\,. 
\end{equation} 
The theoretical value given in~\cite{Pa2001} is different: 
\begin{equation} 
\label{nuthOLD} 
\nu_{\rm th, old}(2{\rm S}_{1/2} - 2{\rm P}_{1/2}) = 14041.57(8) \, { 
\rm MHz}\,, 
\end{equation} 
where the uncertainty has been assigned according to an estimate of the 
magnitude of higher-order contributions which had been unknown. If 
these higher unknown terms had turned out to increase rather than 
decrease the value of the Lamb shift, then a $3.0\,\sigma$ discrepancy 
between theory and experiment could have arisen. In the current paper, 
we review a shift in the theoretical value due to recently calculated 
higher-order two-loop self-energy corrections, and a further shift due 
to radiative-recoil corrections where a certain discrepancy between 
conflicting results has recently been eliminated. 

Bound-state quantum electrodynamic (QED) energy shifts may be expressed 
as combined perturbation series in the parameter $\alpha$, which is the 
fine-structure constant, and in $Z\alpha$, which is a measure of the 
electron-nucleus interaction. Although Lamb-shift measurements in 
atomic hydrogen ($Z=1$) have a long history, it is necessary to perform 
at least one measurement for a different value of $Z$ in order to test 
the $Z$-dependence of the corrections, which are generated by 
bound-state effects and are fundamentally different from the QED of 
free electrons. In He$^+$, we have $Z=2$, and the series in $Z\alpha$ 
converge much slower than for $Z=1$ (atomic hydrogen). 

A more accurate measurement of the Lamb shift in He$^+$ is currently 
being pursued by Hinds and Boshier~\cite{Hipriv}. It is perhaps 
worthwhile to note that He$^+$, as it constitutes an ionic system, 
should also be an attractive candidate for high-precision 
laser-spectroscopic measurements in traps, where the small Doppler 
shifts associated with the slow movement of the trapped ion could 
provide a basis for a further significant reduction in the experimental 
uncertainty. 

%
%
\section{Evaluation of Higher-Order Corrections} 

\begin{figure}[htb] 
\begin{center} 
\begin{minipage}{12cm} 
\centerline{\mbox{\epsfysize=5.0cm\epsffile{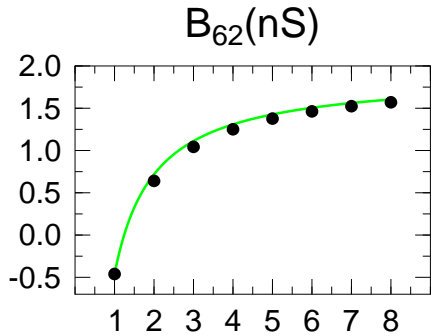}}\hbox to 
0.75in{}} 
\caption{\label{fig1} Dependence of the $B_{62}$-coefficient on the 
principal quantum number $n$ for S states. On the abscissa, we have the 
principal quantum number $n$, and the ordinate axis represents the 
numerical values of $B_{62}(n{\rm S})$ as given in 
equation~(\ref{resb62}). The smooth curve results from a 
three-parameter fit with a model of the form $a + b/n + c/n^2$ (the 
quantities $a$, $b$, and $c$ are fit parameters). A model of this form 
has been shown to lead to a satisfactory representation of the 
$n$-dependence of quantum electrodynamic radiative corrections in many 
cases (see~\cite{Je2003jpa} and references therein). The coefficient 
$B_{62}$ changes sign between $n=1$ and $n=2$.} 
\end{minipage} 
\end{center} 
\end{figure}

A detailed summary of radiative corrections which contribute to the 
Lamb shift of $n=2$ states in He$^+$ has been given in Table~I 
of\cite{vWHoDr2001}. Recently, the understanding of two-loop binding 
corrections to the Lamb shift of S states has been significantly 
enhanced~\cite{Pa2001}. Terms of order 
\begin{equation} 
\left(\frac{\alpha}{\pi}\right)^2\, \frac{(Z\alpha)^6\,m_{\rm e}\,c^2} 
{n^3}\, \ln^2[(Z\alpha)^{-2}] \, B_{62}(n{\rm S}) \quad \mbox{and} 
\quad \left(\frac{\alpha}{\pi}\right)^2\, \frac{(Z\alpha)^6\,m_{\rm e} 
\,c^2}{n^3}\, \ln[(Z\alpha)^{-2}] \, B_{61}(n{\rm S}) 
\end{equation} 
have been evaluated. [Here, $m_{\rm e}$ is the electron mass, $c$ is 
the speed of light, and the energy shift can be converted to 
frequencies via division by the natural unit of action, which is 
Planck's constant $h$. In the following, we use a system of units in 
which $\hbar = \epsilon_0 = c = 1$. We will also use the notations $m$ 
and $m_{\rm e}$ for the electron mass interchangeably~\cite{vWHoDr2001},
and by $M$, we denote the mass of the $\alpha$ particle.] These 
coefficients involve contributions originating from two-loop 
self-energy, combined self-energy vacuum-polarization, and two-loop 
vacuum-polarization diagrams. The final results~\cite{Pa2001} read as 
follows, 
\begin{subequations} 
\label{resb62} 
\begin{eqnarray} 
B_{62}(1{\rm S}) &=& \frac{104}{135} - \frac{16}{9}\,\ln 2 = -0.461
\,891 \,, \\[0.5ex] 
B_{62}(n{\rm S}) &=& B_{62}(1{\rm S}) + \frac{16}{9} \left( \frac34 + 
\frac{1}{4 n^2} - \frac1n - \ln(n) + \Psi(n) + \gamma\right)\,, 
\end{eqnarray} 
\end{subequations} 
where $\gamma = 0.577\,21\dots$ is Euler's constant, and $\psi(z) = 
\Gamma'(z)/\Gamma(z)$. For $n\to\infty$, the coefficient approaches a 
limiting value of 
\begin{equation} 
\lim_{n\to\infty} B_{62}(n{\rm S}) = \frac{4}{135}\, \left[71 + 60 \, ( 
\gamma - \ln2)\right] = 1.897\,603\,. 
\end{equation} 
The numerical value of the $B_{62}$-coefficient depends rather 
significantly on $n$. For $n=2$, we have 
\begin{equation} 
B_{62}(2{\rm S}) = \frac{419}{135} - \frac{32}{9}\,\ln2 = 0.639~180\,. 
\end{equation} 
A graph of $B_{62}(n{\rm S})$ in the range $n = 1,\dots,8$ is given in 
Fig.~\ref{fig1}. 

%
%

Surprisingly large results have been obtained~\cite{Pa2001} for the 
coefficient $B_{61}$, 
\begin{subequations} 
\begin{eqnarray} 
\label{res1S} 
B_{61}(1{\rm S}) &=& \frac{39\,751}{10\,800} + \frac{110}{9}\,\zeta(2) + 
\frac92\, \,\zeta(2)\,\ln 2 - \frac98 \, \zeta(3) \nonumber\\[0.5ex] 
& & - \frac{616}{135}\,\ln 2 + \frac{40}{9} \, \ln^2 2 + \frac43 \, 
N(1{ \rm S}) = 50.344\,005\,, \\[0.5ex] 
B_{61}(n{\rm S}) &=& B_{61}(1{\rm S}) + \frac43 \, \left[ N(n{\rm S}) - 
N(1{\rm S})\right] \\[0.5ex] 
& & + \left( \frac{304}{135} - \frac{32}{9} \, \ln 2\right) \, \left( 
\frac34 + \frac{1}{4 n^2} - \frac1n - \ln(n) + \Psi(n) + \gamma\right) 
\,.\nonumber 
\end{eqnarray} 
\end{subequations} 
The quantity $N(n{\rm S})$ results from a correction to the Bethe 
logarithm induced by a local potential, and has been given for excited 
states in~\cite{Je2003jpa}. In particular, we have $N(2{\rm S}) = 
12.032\,209(1)$ and $B_{61}(2{\rm S}) = 42.447\,669(1)$. This is a 
$15\,\%$ deviation from the corresponding result for 1S~\cite{Pa2001} 
which is indicated here in equation~(\ref{res1S}). 
The $B_{61}$-correction, for 2S, therefore does not quite enhance the 
2S$_{1/2}$-2P$_{1/2}$ interval as much as might be expected from the 
corresponding result for the 1S state [equation~(\ref{res1S})]. 

Additionally, we note that the recently available~\cite{PaJe2003} 
nonlogarithmic two-loop terms of order 
\begin{equation} 
\left(\frac{\alpha}{\pi}\right)^2\, \frac{(Z\alpha)^6\,m_{\rm e}\,c^2} 
{n^3}\, B_{60}(n{\rm S}) 
\end{equation} 
result in a negative energy shift for S states and that the 
$B_{60}$-coefficients are surprisingly large, 
\begin{subequations} 
\label{resb60} 
\begin{eqnarray} 
B_{60}(1{\rm S}) &=& -61.6(3)\pm 15\% 
\label{21} 
\,, \\ 
B_{60}(2{\rm S}) &=& -53.2(3)\pm 15\% 
\label{22} 
\,. 
\end{eqnarray} 
\end{subequations} 
Here, the uncertainty is mainly due to an unknown high-energy 
contribution to this coefficient. 

Recent progress has also been reported with regard to recoil 
corrections of first order in the mass ratio $m_{\rm e}/M$, where $M$ 
is the mass of the atomic nucleus. For hydrogenic systems, a 
nonperturbative evaluation to all orders in $Z\alpha$ has been 
performed in~\cite{ArShYe1995pra}, and the results are in agreement 
with the known terms of the $Z\alpha$-expansion of this effect. These 
are as follows: in the order $(Z\alpha)^5 \, (m_{\rm e}/M) \, m_{\rm e} 
\,c^2$, we have the Salpeter correction. In relative order $(Z\alpha)$, 
we have a correction obtained by Pachucki and Grotch~\cite{PaGr1995}. 
Finally, in relative order $(Z\alpha)^2\,\ln^2[(Z\alpha)^{-2}]$, we 
have a further correction recently obtained by Pachucki and 
Karshenboim, and Melnikov and Yelkhovsky~\cite{PaKa1999,MeYe1999}. The 
sum of these corrections, evaluated for $Z=2$, leads to a satisfactory 
agreement with the numerical data published in~\cite{ArShYe1995pra} for 
the entire range $1 < Z \leq 100$. The nonperturbative (in $Z\alpha$) 
remainder can conservatively be estimated as $3\,{\rm kHz}$, which is 
three times the magnitude of the term of order $(Z\alpha) 
^2\,\ln^2[(Z\alpha)^{-2}]$ (see~\cite{PaKa1999,MeYe1999}). This term is 
{\em not} the main source of the theoretical uncertainty. 

A discrepancy in the analysis of radiative-recoil corrections of order 
$\alpha\,(Z\alpha)^5 \, (m_{\rm e}/M) \, m_{\rm e}\,c^2$ has recently 
been resolved. According to the fully analytic 
calculation~\cite{EiGrSh2001pra}, the coefficient multiplying this term 
has the value $-1.324\,028/n^3$ for S states (here, $n$ is the 
principal quantum number). In~\cite{PaGr1995}, a slightly different 
coefficient of $-1.374/n^3$ had been obtained, whereas 
in~\cite{BhGr1985,BhGr1987aop,BhGr1987}, a coefficient of $-1.988(4)$ 
has been given. Assuming the correctness of the most recent 
evaluation~\cite{EiGrSh2001pra}, the result for radiative-recoil 
corrections of order $\alpha\,(Z\alpha)^5 \, (m_{\rm e}/M) \, m_{\rm e} 
\,c^2$ should be taken as $-0.014\,{\rm MHz}$ rather than $-0.035\,{\rm 
MHz}$. This concerns the entry termed ``$\alpha\,(Z\alpha)^5\,m/M$'' in 
Table I of~\cite{vWHoDr2001} which, in units of $\alpha\,m_{\rm e} 
\,c^2$, is of relative order $(Z\alpha)^5\,m_{\rm e}/M$. 

Finally, we note that the uncertainty due to the nonperturbative (in 
$Z\alpha$) one-loop self-energy remainder $G_{\rm SE}(Z\alpha)$ has 
recently been eliminated~\cite{JeMoSo2001pra}: the numerical result 
reads $G_{\rm SE}(2{\rm S}_{1/2}, 2\alpha) = -30.644\,66(5)$. For the 
2S state of He$^+$, the following energy shifts result from the results 
reported in equations (\ref{resb62})--(\ref{resb60}): 
\begin{subequations} 
\label{results} 
\begin{eqnarray} 
B_{62}(2{\rm S}) &\to& +0.037\,{\rm MHz}\\ 
& & \mbox{(entry termed $\alpha\,(Z\alpha)^6\,\ln^2(Z\alpha)^{-2}$ in 
Table I of~\cite{vWHoDr2001})} \,, \nonumber\\ 
B_{61}(2{\rm S}) &\to& +0.289\,{\rm MHz}\\ 
& & \mbox{(entry termed $\alpha\,(Z\alpha)^6\,\ln(Z\alpha)^{-2}$ in 
Table I of~\cite{vWHoDr2001})} \,, \nonumber\\ 
B_{60}(2{\rm S}) &\to& (-0.043 \pm 0.006)\,{\rm MHz}\\ 
& & \mbox{(entry termed $\alpha\,(Z\alpha)^6$ in Table I of~ 
\cite{vWHoDr2001})} \,, \nonumber\\ 
\mbox{radiative recoil} &\to& -0.014\,{\rm MHz}\\ 
& & \mbox{(entry termed $\alpha\,(Z\alpha)^5\,m/M$ in Table I of~ 
\cite{vWHoDr2001})} \,, \nonumber\\ 
G_{\rm SE}(2{\rm S}_{1/2}, 2\alpha) &\to& -10.624\,{\rm MHz}\\ 
& & \mbox{(entry termed $(Z\alpha)^6\,G_{\rm SE}(Z\alpha)$ in Table I 
of~\cite{vWHoDr2001})} \,, \nonumber 
\end{eqnarray} 
\end{subequations} 
For completeness, we should also note slight inconsistencies in the 
notation of the corrections in~\cite{vWHoDr2001}. 
First, in Table I of~\cite{vWHoDr2001}, all corrections are given in 
units of $\alpha\,m_{\rm e}\,c^2$. The Salpeter correction has 
inadvertently been termed ``$(Z\alpha)^5\,m/M$'' whereas in units of 
$\alpha\,m_{\rm e}\,c^2$, it should be termed ``$Z\,(Z\alpha) 
^4\,m/M$.'' Correspondingly, the correction denoted ``$(Z\alpha) 
^6\,m_{\rm e}\,c^2$'' should actually be termed ``$Z\,(Z\alpha) 
^5\,m/M$'', and the correction denoted as ``$(Z\alpha)^7\,\ln(Z\alpha) 
\,m/M$'' is of course a correction of (relative) order ``$Z\,(Z\alpha) 
^6\,\ln(Z\alpha)\,m/M$.'' 

We now discuss the main source of the theoretical uncertainty. There 
exists an essentially unknown three-loop binding correction of order 
\begin{equation} 
\left(\frac{\alpha}{\pi}\right)^3\, \frac{(Z\alpha)^5\,m_{\rm e}\,c^2} 
{n^3}\, C_{50}(n{\rm S})\,, 
\end{equation} 
where we note that the $C_{50}$-coefficient, in analogy to the two-loop 
binding correction $B_{50} \approx -21.5562(31)$, might be numerically 
large. With the estimate $|C_{50}| \approx 30$ for the unknown term, 
the corresponding additional uncertainty for the 2S state is 
$0.008\,{\rm MHz}$. The nuclear-size contribution to the uncertainty is 
$0.010\,{\rm MHz}$~\cite{BoRi1978,vWHoDr2001} (this is assuming a 
nuclear radius of $1.673(1)\,{\rm fm}$). There is an additional 
$0.015\,{\rm MHz}$~uncertainty due to unknown higher-order two-loop 
corrections for 2P states. We obtain the following theoretical error 
budget for the theoretical value of the Lamb shift in He$^+$: 
\begin{itemize} 
\item unknown high-energy part of $B_{60}$: $6\,{\rm kHz}$. 
\item unknown higher-order two-loop effects for P states: $15\,{\rm 
kHz}$. 
\item unknown three-loop binding correction $C_{50}$: $8\,{\rm kHz}$. 
\item nonperturbative remainder of nuclear recoil: $3\,{\rm kHz}$. 
\item uncertainty in the nuclear charge radius: $10\,{\rm kHz}$. 
\end{itemize} 
The $ \alpha$ particle charge radius has been determined to high 
accuracy in~\cite{BoRi1978}, using spectroscopic experimental data from 
the system $\mu^-$--He$^{2+}$. However, the accuracy of the resulting 
charge radius of $1.673(1)\,{\rm fm}$ has been questioned (see 
e.g.~\cite{Co1982pra,BrZa1990}). The result obtained in combining 
scattering experiments is $1.674(12)\,{\rm fm}$~\cite{SiCaWh1976} and 
has a considerably larger error. The above entry in the error budget 
should therefore rather be interpreted as a lower limit of this 
problematic contribution. Because theoretical errors cannot be expected 
to follow a normal distribution like experimental errors, there exists 
no universally adopted procedure for the determination of a total 
theoretical error in a situation where several unknown terms contribute 
to an error budget. In the current work, we choose to (conservatively) 
add the above errors in order to obtain a total theoretical uncertainty 
of $42\,{\rm kHz}$. To complete the discussion, we mention that (i) 
under the assumption of normally distributed theoretical errors, the 
total theoretical uncertainty would be reduced to $20\,{\rm kHz}$, and 
(ii) under the assumption of a $0.012\,{\rm fm}$ uncertainty in the 
nuclear radius~\cite{SiCaWh1976}, the theoretical error due to the 
$\alpha$ particle charge radius alone would be $126\,{\rm kHz}$. 

%
%
\section{Conclusions} 

We have re-evaluated theoretical predictions for the $2{\rm S}$--$2{\rm 
P}_{1/2}$ Lamb shift in He$^+$. Taking into account recently evaluated 
corrections to the entries in Table I of~\cite{vWHoDr2001} (see 
equation~(\ref{results})), we obtain the new theoretical value of 
\begin{subequations} 
\begin{equation} 
\label{nuth2S2P12} 
\nu_{\rm th}(2{\rm S}_{1/2} - 2{\rm P}_{1/2}) = 14041.474(42)(126) \, { 
\rm MHz}\,, 
\end{equation} 
where the first error is due to the error budget as discussed above, 
and the second error results due to the nuclear size alone, if we 
assign a larger nuclear charge radius uncertainty of $0.012\,{\rm 
fm}$~\cite{SiCaWh1976} instead of $0.001\,{\rm fm}$~\cite{BoRi1978}. 
For the 2P$_{3/2}$--2S$_{1/2}$-interval, the new theoretical prediction 
is 
\begin{equation} 
\label{nuth2P322S} 
\nu_{\rm th}(2{\rm P}_{3/2} - 2{\rm S}_{1/2}) = 161542.029(42)(126) \, 
{ \rm MHz}\,, 
\end{equation} 
\end{subequations} 
with the same error budgets as for $2{\rm S}_{1/2} - 2{\rm P}_{1/2}$. 

There remains a small discrepancy between the ``new'' theoretical value 
of $14041.474(42) \, {\rm MHz}$ (equation~(\ref{nuth2S2P12})) and the 
experimental result of $14041.13(17) \, {\rm MHz}$ given in 
equation~(\ref{nuexp}). However, this discrepancy is less severe than 
the difference of the experimental value~(\ref{nuexp}) and the ``old'' 
theoretical value of $14041.57(8) \, {\rm MHz}$ (see equation 
(\ref{nuthOLD})). 

The sum of the experimental ($=0.17\,{\rm MHz}$) and the theoretical 
uncertainty ($=0.04\,{\rm MHz}$) is $\sigma = 0.21\,{\rm MHz}$. This 
should be compared to the difference $\delta = 0.34\,{\rm MHz}$ between 
the theoretical and experimental ``expectation values'' 
(\ref{nuth2S2P12}) and (\ref{nuexp}). The difference $\delta$ 
corresponds to $1.6\,\sigma$. If we assign the larger error of 
$0.012\,{\rm fm}$ to the nuclear radius, then the discrepancy shrinks 
to $1.2\,\sigma$. 

Even though a difference of $1.2\,\sigma$ to $1.6\,\sigma$ between 
theory and experiment remains somewhat unsettling, it is not large 
enough to be regarded as statistically significant.  Even at this 
level, it still represents a better test of the unexpectedly large 
$B_{50}$ two-loop binding correction to the Lamb shift (--1.339 MHz) 
than that provided by the corresponding measurement in hydrogen.  In 
addition, the anisotropy method provides a completely independent 
method of measuring the Lamb shift that is not limited by the large 
level width of the $2{\rm P}$ state(s), and it is the only method that 
is capable of comparing two different atomic or ionic species (H and 
He$^+$, c.f.\ Ref.\ \cite{Wijn98}) within the same apparatus. 

It may also be beneficial, in the near future, to re-analyze the 
results of the most recent measurement~\cite{vWHoDr2001} in terms of 
QED corrections to the $2{\rm S}$--$2{\rm P}$ transition matrix 
elements which are currently under study~\cite{Sa2003priv}. In the 
current paper, we restrict the discussion to the {\em theoretical} 
predictions for the absolute frequency of the $2{\rm S}$--$2{\rm P}$ 
transitions. Further clarification of this and related questions will 
certainly benefit from a  reduction of the experimental 
uncertainty of the Lamb-shift measurement to $\pm$100 kHz or better,
as is currently being 
pursued by Hinds and Boshier~\cite{Hipriv}. Finally, we stress 
that all theoretical predictions for transitions in He$^+$ rely on the 
single available accurate determination of the charge radius of the 
$\alpha$-particle as described in~\cite{BoRi1978}. There is still a 
lack of accurate independent verifications of this determination of the 
charge radius. In the end, the answer to a number of questions related 
to Lamb-shift measurements in atomic hydrogen and He$^+$ might be as 
easy as a slight deviation of the proton and $\alpha$ particle charge 
radii from their currently accepted values. However, other, more subtle 
explanations cannot be ruled out at present, either. 

%
%
\section*{Acknowledgements} 

The authors acknowledge helpful conversation with P. J. Mohr and J. 
Sapirstein. We are grateful for SHARCNET support extended to the 
University of Windsor during the organization of a workshop related to 
topical questions in quantum electrodynamics, on the occasion of which 
this research note has been completed.  Research support by the Natural 
Sciences and Engineering Research Council of Canada is gratefully 
acknowledged.

\end{document}